\documentclass[10pt]{article}%
\usepackage{amssymb}
\usepackage{amsfonts}
\usepackage{epsf}
\usepackage{amsmath}
\usepackage{graphicx}%
\def\gtrsim{\mathrel{\raise.3ex\hbox{$>$\kern-.75em\lower1ex\hbox{$\sim$}}}}

\def\-{\hphantom{-}}

\def\s2{\frac{1}{\sqrt2}}

\def\beq{\begin{equation}}
\def\eeq{\end{equation}}
\def\beqa{\begin{eqnarray}}
\def\eeqa{\end{eqnarray}}
\def\bea{\begin{eqnarray}}
\def\eea{\end{eqnarray}}
\def\be{\begin{equation}}
\def\ee{\end{equation}}

\def\mg{m_{3/2}}
\def\mg2{m^2_{3/2}}

\def\Dsl{\,\raise.15ex\hbox{/}\mkern-13.5mu D}

\begin{document}
\pagestyle{plain}


\makeatletter
\makeatother
\pagestyle{empty}
\rightline{ IFT-UAM/CSIC-06-06} \rightline{ OUTP-07-01-P}

\begin{center}
{\LARGE {Supersymmetric Higgs and Radiative Electroweak Breaking \\[20mm]}
}{\large {\ L.~E.~Ib\'{a}\~{n}ez$^{a}$ and G.G.Ross$^{b}$ \\[6mm]}
}{\small {$^{a}$ Departamento de F\'{\i}sica Te\'{o}rica C-XI and
\\[0pt]Instituto de F\'{\i}sica Te\'{o}rica C-XVI\\[0pt]Universidad
Aut\'{o}noma de Madrid\\[0pt]Cantoblanco, 28049 Madrid, Spain.\\[8mm]} {$^{b}$
Rudolf Peierls Centre for Theoretical Physics, \\[0pt]University of
Oxford,\\[0pt]Oxford OX13NP.\\[20mm]} }

{\small \textbf{Abstract} \\[10mm]}
\end{center}

{\small We review the mechanism of radiative electroweak symmetry breaking
taking place in SUSY versions of the standard model. We further discuss
different proposals for the origin of SUSY-breaking and the corresponding
induced SUSY-breaking soft terms. Several proposals for the understanding of
the little hierarchy problem are critically discussed. }

\vspace{5.0cm}



\setcounter{page}{1} \pagestyle{plain}
\renewcommand{\thefootnote}{\arabic{footnote}}

\section{Introduction}

The driving force in seeking what lies beyond the Standard Model has been the
desire to understand electroweak symmetry breaking and the generation of mass.
The Standard Model parameterises this breaking via a stage of spontaneous
symmetry breaking generated by a component of the "Higgs\textquotedblright%
\ scalar electroweak doublet field. However it leaves many fundamental
questions unanswered. It does not explain what the scale of symmetry breaking
should be and provides no underlying reason for the complicated gauge and
matter multiplet structure. In addition it requires a large number of
parameters to be specified to determine the various interactions needed by the model.

Attempts to explain the structure of the Standard Model usually postulate new
physics at a scale above the electroweak scale. The archetypical example of
this is Grand Unification or heterotic superstring unification in which the
Standard Model gauge group $SU(3)\times SU(2)\times U(1)$ is a subgroup of a
simple gauge group such as $SU(5)$, $SO(10)$ or $E_{6},$ broken at a very high
scale close to the Planck scale. By unifying the strong, weak and
electromagnetic interactions these theories simplify the gauge and multiplet
structure and relate many of the parameters of the Standard Model. However
Grand Unified models still leave unanswered the fundamental question why the
scale of electroweak breaking should be so much less than the Grand Unified
scale or the string scale. The problem is particularly acute because radiative
corrections to the Higgs sector drive this scale close to the underlying
unified scale. To see this consider the potential for the Standard Model
Higgs, $h$, given by%

\begin{equation}
V={\frac{1}{2}}m^{2}h^{2}+{\frac{1}{4}}\lambda h^{4}\ , \label{one}%
\end{equation}
At one loop order the mass has quadratically-divergent contributions
\cite{veltman}. Treating the Standard Model as an effective field theory valid
at scales below $\Lambda_{\mathrm{SM}}$ this contribution is given by
\begin{equation}
\delta_{q}m^{2}={\frac{3}{64\pi^{2}}}(3g^{2}+g^{\prime2}+8\lambda-8\lambda
_{t}^{2})\Lambda_{\mathrm{SM}}^{2}\ , \label{hierarchysm}%
\end{equation}
where $g,g^{\prime},\lambda$ and $\lambda_{t}$ are the $SU(2)\times U(1)_{Y}$
gauge couplings, the quartic Higgs coupling and the top Yukawa coupling,
respectively. The requirement of no fine-tuning between the above contribution
and the tree-level value of $m^{2}$ sets an upper bound on $\Lambda
_{\mathrm{SM}}$. For a Higgs mass in the range $m_{h}=115-200$ GeV,
\begin{equation}
\left\vert {\frac{\delta_{q}m^{2}}{m^{2}}}\right\vert \leq10\ \Rightarrow
\ \Lambda_{\mathrm{SM}}\leqslant2-3\ \mathrm{TeV}\ . \label{eftcutoffsm}%
\end{equation}
This is to be compared to the Grand Unified scale or string scale some
thirteen or more orders of magnitude higher, which is the expected cutoff in
the original unified extensions of the Standard Model! This is the hierarchy
problem. \ A separation of these scales is possible if the electroweak
breaking scale is protected from large radiative corrections by a
(spontaneously broken) symmetry. While some limited protection is possible if
the Higgs is a pseudo-Goldstone boson (see below) the only symmetry capable of
providing protection from a very high scale is supersymmetry (SUSY). This
requires that the Standard Model states be accompanied by supersymmetric
partners, the squarks and sleptons, the Higgsinos and the gauginos which must
get a mass when supersymmetry is broken, the mass scale for these states being
limited to be $\leqslant\Lambda_{\mathrm{SM}}$ \textit{i.e.} roughly within an
order of magnitude of the electroweak breaking scale by the need to solve the
hierarchy problem.

The breaking of supersymmetry initially seemed to be the stumbling block to
implementing the supersymmetric solution to the hierarchy problem because it
was known that if supersymmetry breaking is directly coupled to the Standard
Model states then the mas of the superpartners obey a sum rule \cite{fgp} such
that a component of each of the squarks and sleptons would be lighter than its
quark and lepton partners. Since we have not observed such light scalar states
carrying Standard Model quantum numbers this is not a viable possibility.
However if supersymmetry should be broken in an "hidden" sector with no direct
couplings to the SM states then the sum rule can be evaded with both
components of the squarks and sleptons heavier than their fermion partners.
This is because SUSY\ breaking is only communicated to the visible sector by
radiative and/or gravitational corrections involving
"messenger\textquotedblright\ states and these effects do not obey the sum
rule\cite{Ibanez:1982fr},\cite{Inoue:1982pi},\cite{Alvarez-Gaume:1983gj}.
Initial attempts to explain the magnitude of the hierarchy employed additional
symmetries to force the supersymmetry breaking, driven by Yukawa couplings
involving messenger states, to be much smaller than the underlying
supersymmetry breaking in the hidden sector\cite{Ellis:1982fc}. Subsequently
it was realised that gravitational couplings (SUGRA messengers) present in
supergravity theories (theories in which the supersymmetry is made a local
symmetry) also provides a coupling between the hidden and visible sectors that
makes the supersymmetric scalar partners of the quarks and leptons heavier
than their fermion partners \cite{gm} . Given the inevitable presence of such
supergravity corrections this source of supersymmetry breaking in the hidden
sector has enjoyed great popularity. Due to the weakness of gravitational
coupling, this \textquotedblleft SUGRA\textquotedblright\ origin for visible
sector supersymmetry breaking also is consistent with a high scale of hidden
sector scale of supersymmetry breaking, of $O(10^{10}GeV)$.

In constructing phenomenologically viable models of supersymmetry breaking it
is essential not to introduce flavour changing neutral current (FCNC)
processes at a level much greater than that generated in the Standard Model.
This constraint has proved to be extremely restrictive because radiative
contributions to flavour changing processes have their supersymmetric analogue
in supersymmetric extension of the Standard Model. The simplest way to satisfy
the flavour changing constraints is if the squarks and sleptons of the three
families are degenerate at the SUSY breaking messenger scale. To achieve this
several classes of theories have been constructed in which the coupling
between the hidden and visible sectors is due to gravity or to gauge
interactions. In the case of gravity, \textquotedblleft
SUGRA\textquotedblright\ models, it is not guaranteed that the soft masses are
family independent but there are, for example, specific superstring models
with this property. For the case of gauge theories, \textquotedblleft Gauge
Mediated models\textquotedblright, the communication of SUSY\ breaking from
the hidden to the visible sector is by a gauge interaction \cite{gaugem} which
is the same for each family. In these schemes the underlying hidden sector
scale of supersymmetry breaking can be much smaller (see below).

While supersymmetry offers a solution to the gauge hierarchy problem it
introduces new questions. The introduction of scalar partners to the quarks
and leptons now means that they, like the Higgs scalar, can acquire vacuum
expectation values triggering spontaneous symmetry breaking. This raises the
question why should the gauge symmetry group $SU(3)\times SU(2)\times U(1)$ be
broken to $SU(3)\times U(1)$ rather than some other subgroup? In theories with
Grand Unification or string unification the need to explain why it is the
electroweak group that is spontaneously broken becomes even more pressing
because each gauge group factor of the Standard Model is on the same footing.

It was in the context of hidden sector SUSY\ breaking that radiative
electroweak breaking was discovered and shown to provide an elegant answer to
why it is the electroweak group that is broken and why the breaking scale
should be low. Whatever the messenger sector, an initially degenerate spectrum
of scalar states at the messenger scale will be non-degenerate at lower scales
due to radiative corrections. It is this that naturally leads to radiative
electroweak breaking because the radiative corrections can drive the mass
squared of a scalar state negative at a scale below the messenger mass scale,
triggering spontaneous symmetry breaking at a scale close to the supersymmetry
breaking scale in the visible sector. The latter must be low if the hierarchy
problem is to be solved. Radiative corrections due to the couplings in the
superpotential give a negative contribution to scalar masses squared while
those due to gauge couplings give a positive contribution. This means that the
direction of spontaneous symmetry breaking is determined by the couplings of
the theory. As we shall discuss this provides a natural explanation for the
breaking of the Standard Model group to be in the direction $SU(3)\times
SU(2)\times U(1)\rightarrow SU(3)\times U(1)_{EM}.$

The price for the supersymmetric solution to the hierarchy problem is that
there is a spectrum of supersymmetric states in the visible sector with mass
of order the electroweak breaking scale. The fact none has been observed is
already somewhat surprising because it reintroduces the need for some fine
tuning of parameters to separate the electroweak and supersymmetry breaking
scales; this is the \textquotedblleft little hierarchy
problem\textquotedblright.\ However there \textit{is} some circumstantial
evidence for the presence of the light supersymmetric partners of the Standard
Model states. This comes from the fact that only with their inclusion do the
gauge couplings of the Standard Model unify\cite{Ibanez:1981yh}%
,\cite{Dimopoulos:1981yj} suggesting an underlying stage of unification of the
forces. The fact that they unify to a very high degree of accuracy, better
than $1\%$\cite{Ghilencea:2001qq}, is one of the main reasons why
supersymmetry is widely considered to be the most likely extension of the
Standard model.

In this article we will review the mechanism giving rise to radiative
electroweak breaking in supersymmetric theories. In Section 2 we briefly
introduce the structure of the minimal supersymmetric extension of the
Standard Model (MSSM). Central to the breaking of the Standard Model gauge
group are the soft supersymmetry breaking masses and we discuss the various
proposals for the soft SUSY\ breaking terms both in a field theory and a
string theory context. In Section 3 we discuss the structure of radiative
corrections driving radiative electroweak breaking using the technique of
renormalisation group equations and show how it explains the pattern of
symmetry breaking observed in the Standard Model. In Section 4 we present a
discussion of the "little hierarchy problem", comparing the case of
supersymmetric solutions to various alternatives that have been suggested. In
Section 5 we present a summary and discuss the implications for future
experimental searches for the Higgs boson(s) and the supersymmetric partners
of Standard Model states.

\section{Supersymmetric extension of the Standard Model - the MSSM}

The simplest viable supersymmetric extension of the Standard Model involves a
single ($N=1$) supersymmetry generator commuting with the Standard Model gauge
group so that the symmetry group is the direct product $SU(3)\times
SU(2)\times U(1)\times(N=1$ $SUSY).$ The gauge bosons are assigned to vector
supermultiplets together with their fermion \textquotedblleft
gaugino\textquotedblright\ superpartners. The quarks and leptons are assigned
to chiral supermultiplets together with their scalar squark and slepton
superpartners. Finally it is necessary to introduce two $SU(2)$ doublets of
Higgs scalars (carrying equal but opposite weak hypercharge) making up
additional chiral supermultiplets together with their fermion Higgsino
superpartners. In the minimal version of the theory the lightest
supersymmetric state (the LSP) is stable and is a candidate for dark matter.

\subsection{The couplings}

The gauge interactions of the states of the theory are now determined. Due to
the direct product structure of the gauge symmetry group the new
supersymmetric states carry the same gauge quantum numbers as their Standard
Model partners and so they have the same couplings to the gauge bosons.
Operation by the supersymmetry generator induces new couplings involving the
gaugino partner of the gauge bosons and we will not reproduce them
here\footnote{The detailed form of the interactions in supersymmetric theories
have been extensively discssed in numerous articles and books. For example see
\cite{Martin:1997ns} and references therein.}.

The Yukawa couplings and associated quartic scalar couplings come from the
superpotential. To reproduce the couplings of the Standard Model requires the
following superpotential, $P$%
\begin{equation}
P=h_{ijk}L_{i}H_{1j}{E}_{k}\ +\ h{^{\prime}}_{ijk}Q_{i}H_{1j}{D}%
_{k}\ +\ h{^{\prime\prime}}_{ijk}Q_{i}H_{2j}{U}_{k} \label{superp}%
\end{equation}
where $L$ and $E$ ($Q$ and $U,D$ ) are the (left-handed) lepton doublet and
antilepton singlet (quark doublet and antiquark singlet) chiral superfields
respectively and $H_{1,2}$ are (left-handed) Higgs superfields. The family
indices, $i$, $j$ and $k$ are summed over the three families. The
supersymmetric couplings correspond to the F terms of the superpotential $P$.
These give both Yukawa couplings and pure scalar couplings. For example, the
Yukawa couplings following from the first term of eq.(\ref{superp})\ are
\begin{equation}
L_{Yukawa}\ =\ h_{ijk}(L_{i}H_{1j}{E}_{k}+{\tilde{L}}_{i}{\tilde{H}}_{1j}%
E_{k}+L_{i}{\tilde{H}}_{1j}{\tilde{E}}_{k}) \label{yuki}%
\end{equation}
where we denote by a supertwiddle the supersymmetric partners to the quarks,
leptons and Higgs bosons, namely the squarks, sleptons and Higgsinos. The
first term is the usual term in the Standard Model needed to give charged
leptons a mass. The new couplings associated with the supersymmetric states
related to the first term by the operation of the supersymmetry generator are
given by the second and third terms. The scalar couplings associated with
eq.\ref{superp}\ are%

\begin{equation}
L_{scalar}\ =\ \sum_{i,j}\mid{\frac{{\partial^{2}P}}{{\partial\phi_{i}\phi
_{j}}}}\mid^{2} \label{poti}%
\end{equation}
where $\phi_{i}$ are chiral superfields and after differentiation of the
superpotential only the scalar components of the remaining chiral superfields
are kept. The full set of Feynman rules resulting from the superpotential of
eq.\ref{superp} may be found e.g. in ref. \cite{Martin:1997ns}.

There is one further coupling needed to complete the couplings of the minimal
supersymmetric version of the standard model. In order to generate a mass for
the Higgsinos associated with the Higgs doublets $H_{1,2}$ it is necessary to
add a term to the superpotential given by%

\begin{equation}
P{\ ^{\prime}}\ =\ \mu H_{1}H_{2} \label{mumu}%
\end{equation}
In addition to giving a mass $\mu$ to the Higgsinos, this term plays an
important role in determining the Higgs scalar potential and the pattern of
electroweak symmetry breaking. As we will discuss in more detail in section 3
the (SUSY-breaking) scalar term following from eq.(\ref{mumu})\ aligns the
vacuum expectation values (vevs) of the two Higgs fields so that the photon is
left massless, obviously a crucial ingredient for a viable theory.

We note that this term is the only one involving a coupling with dimensions of
mass. If the theory is to avoid the hierarchy problem $\mu$ must be small, of
order the electroweak breaking scale, for the Higgs scalars also get a
contribution $\mu^{2}$ to their mass squared. Thus any complete explanation of
the electroweak breaking scale must also explain the origin of $\mu$.

\subsection{Models of soft SUSY\ breaking}

In order to complete the model it is necessary to specify how supersymmetry is
broken. As discussed above the source of supersymmetry breaking has to be in a
hidden sector communicated to the visible sector by a messenger sector.
Provided the supersymmetry breaking is spontaneous the SUSY breaking terms in
the visible sector will be \textquotedblleft soft\textquotedblright\ with
dimension less \ than or equal to 3 in the Lagrangian. This ensures that the
underlying SUSY will still control the hierarchy problem because such soft
terms do not affect the ultraviolet behaviour of the theory. The possible
terms must respect the gauge symmetry of the Standard Model\ (SM). The
resultant effective low energy field theory Lagrangian below the SUSY breaking
scale has the general form
\begin{align}
L_{g}\  &  =\ \frac{1}{2}\sum_{a}M_{a}\ \lambda_{a}\lambda_{a}\ +\ h.c.\\
L_{m^{2}}\  &  =\ -m_{H_{d}}^{2}|H_{d}|^{2}-m_{H_{u}}^{2}|H_{u}|^{2}%
-m_{Q_{ij}}^{2}Q_{i}Q_{j}^{\ast}-m_{U_{ij}}^{2}U_{i}U_{j}^{\ast}\nonumber\\
&  -m_{D_{ij}}^{2}D_{i}D_{j}^{\ast}-m_{L_{ij}}^{2}L_{i}L_{j}^{\ast}-m_{E_{ij}%
}^{2}E_{i}E_{j}^{\ast}\nonumber\\
L_{A,B}\  &  =\ -\ A_{ij}^{U}Q_{i}U_{j}H_{u}\ -\ A_{ij}^{D}Q_{i}D_{j}%
H_{d}\ -\ A_{ij}^{L}L_{i}E_{j}H_{d}\ -\ B\ H_{d}H_{u}\ +\ h.c.\nonumber
\end{align}
in standard notation. This has many (107) free parameters beyond the SM.
However, as mentioned above, there are strong constraints on the form these
terms can take due to the need to avoid large flavour changing processes. For
example the SM $\Delta S=2$ amplitude which generates the $K^{0}-\overline
{K}^{0}$ mass difference has a box diagram contribution involving two
$W-$boson propagators and two fermion propagators involving $u,$ $c$ and $\ t$
quarks. The $GIM$ mechanism forces a cancellation between these contributions
and the resultant contribution is dominated by the $u$ and $c$ contributions
and is proportional to $(m_{c}^{2}-m_{u}^{2})/M_{W}^{2}.$ To suppress the
$\Delta S=2$ contribution below the experimental limit requires $(m_{c}%
^{2}-m_{u}^{2})/M_{W}^{2}\leqslant10^{-4}$ which is satisfied for a charm
quark mass of $O(1GeV)$ as observed.

The analogous contribution from the Standard Model supersymmetric partners
involves two \textquotedblleft Wino\textquotedblright,\ $\widetilde{W},$
fermion propagators, the partners of the $W-$bosons, and two scalar
propagators involving the $\widetilde{u},$ $\widetilde{c}$ and $\ \widetilde
{t}$ squarks, the scalar partners of the $u,$ $c$ and $\ t$ quarks . For Winos
heavier than squarks this contribution is proportional to $(m_{\widetilde{c}%
}^{2}-m_{\widetilde{u}}^{2})/M_{\widetilde{W}}^{2}.$ To suppress the new
SUSY\ contribution below the experimental limit requires\cite{Gabbiani:1996hi}
$(m_{\widetilde{c}}^{2}-m_{\widetilde{u}}^{2})/M_{\widetilde{W}}^{2}%
\leqslant10^{-4}.$ For Wino masses less than $1TeV,$ as is required by the
SUSY\ solution to the hierarchy problem,$\ $implies $m_{\widetilde{c}}%
^{2}-m_{\widetilde{u}}^{2}\lesssim\ 10GeV^{2}$ a surprising result given that
the non-observation of squarks requires that they are much heavier than the
W-boson. This near-degeneracy for the squarks of the first two families is
made even more stringent because, unlike the gluon contribution in the
Standard Model, the gluino contribution in its supersymmetric extension also
changes strangeness and contributes to the box diagram, giving a contribution
enhanced by the strong coupling involved. Given this fact, viable methods of
supersymmetry breaking must explain why the squarks are nearly degenerate.

In specific scenarios of SUSY-breaking many or all the couplings are diagonal
in family space and there are also unification constraints so that the number
of parameters is drastically reduced. This depends on the origin of
SUSY-breaking is. Here we will briefly review the most popular ideas for the
mediation of SUSY-breaking.

\textit{Supergravity mediation}\cite{gm}

This is a well motivated option, since gravity is there anyhow and combined
with SUSY gives rise to supergravity. Asuming there is some hidden
SUSY-breaking sector in the theory, the spin=3/2 partner of the graviton, the
gravitino, gets a mass $m_{3/2}$ of order $<F>/M_{P}$, $<F>\not =0$ being some
SUSY-breaking scalar auxiliary field. One of the generic features of
supergravity is that gauge couplings as well as kinetic terms are field
dependent. The Lagrangian is defined by three classes of functions: the gauge
kinetic functions $f_{a}(\phi_{i})$ (one per group), the Kahler potential
$K(\phi,\phi^{\ast})$ and the superpotential $W$. Soft terms may be explicitly
evaluated in terms of those functions, the gravitino mass, and the values of
the auxiliary fields $F$ breaking SUSY. For example, if the Kahler metrics of
the SM scalar fields are diagonal one can write
\begin{align}
{M_{a}}  &  =\frac{1}{2\mathrm{Re\,}{\ f_{a}}}{F^{M}}\partial_{M}{f_{a}%
}\ ,\nonumber\\
{m_{I}^{2}}  &  ={m_{3/2}^{2}}-\sum_{M,N}{\bar{F}^{\bar{M}}F^{N}}%
\partial_{\bar{M}}\partial_{N}\log({\tilde{K}_{I\bar{I}}})\ ,\label{softt}\\
{A_{IJL}}  &  ={F^{M}}[{\ K_{M}}+\partial_{M}\log(Y_{IJL})-\partial_{M}%
\log({\tilde{K}_{I\bar{I}}\tilde{K}_{J\bar{J}}\tilde{K}_{L\bar{L}}%
)}]\ .\nonumber
\end{align}
where ${\tilde{K}_{II}}$ are the kinetic terms of the SM matter fields. If the
vevs of the auxilary fields are of order $<F^{N}>\propto(10^{11}GeV)^{2} $
then the gravitino mass and the soft terms will be of order the electroweak
scale, $O(10^{2}GeV)$. The simplest scheme of this type is that of the minimal
supergravity model (mSUGRA) which assumes universality of gauge kinetic
functions $f_{a}=f$, minimal matter kinetic terms ${\tilde{K}}_{II}%
=\delta_{II}$ and constant Yukawa couplings. Under these circumstances the
scalar masses are family independent as required by FCNC and there are only 4
SUSY-breaking parameters, $M,$ the common gaugino mass, $m$, the common scalar
mass, $A$, the common coefficient of the soft trilinear terms and $B$ the
coefficient of the soft bilinear term. This scenario is atractive because of
its simplicity and has been very much analyzed in the last twenty years. In
spite of increasing experimental constraints, the mSUGRA secenario is
consistent with radiative EW symmetry breaking and for reduced regions of
parameter space can accomodate the appropriate amount of dark matter in the
form of the relic SUSY LSP(see ref.
\cite{Ellis:2005tu,Allanach:2006jc,Ellis:2006ix,deAustri:2006pe,Ellis:2006nx}
for recent analyses). As we will discuss below, specific gravity mediation
scenarios naturally appear also in the context of string theory, in which the
functions $f_{a},{\tilde{K}},K$ and $W$ may be computed in simple compactifications.

\textit{Gauge mediation}\cite{gaugem,gr}

In this scheme the leading mediators of SUSY-breaking are the SM gauge bosons
and their gaugino partners. In the simplest `minimal' GMSB model \cite{gr} one
has a singlet chiral super field $X$ with $<X>\not =0$ and $<F_{X}>\not =0$,
breaking SUSY. It couples to new, heavy, vector-like fields with SM quantum
numbers which make up complete $SU(5)$ representations (e.g., $(5+{\bar{5}}%
)$). Then SM gaugino masses appear at one loop and scalar masses at two loops
yielding e.g.
\begin{equation}
{M_{a}}\ \simeq\ \frac{{\alpha_{a}}}{4\pi}\left(  \frac{{F_{X}}}{<{X}%
>}\right)  \ ;\ {m_{\tilde{q}}^{2}}\ \simeq\ \left(  \frac{{F_{X}}}{<{X}%
>}\right)  ^{2}\sum_{a}\left(  C_{a}(\frac{{\alpha_{a}}}{4\pi})^{2}\right)
\end{equation}
with $C_{a}$ the quadratic Casimirs of SM groups. This means that in order to
get soft terms of order $M_{W}$ one should have $F_{X}/<X>\simeq100$ TeV. In
this case the LSP is the gravitino ($m_{3/2}\simeq10-100$ eV) and typically
the next-to-LSP(NLSP) is a neutralino or a charged slepton. The typical
experimental signatures for this scheme are processes with missing energy plus
a photon or lepton (the NLSP may decay inside the detector). One serious
problem of the simplest GMSB scenarios is the difficulty in getting an
appropriately large B and/or $\mu$ parameters. On the other hand its great
virtue is that FCNC are very small, since gauge transmission is flavor-blind
and renormalization effects are small.

\textit{Anomaly mediation}\cite{anomm}

This mechanism is a variety of gravity mediation. It was observed that in the
presence of SUSY-breaking in supergravity, there is always a class of one-loop
soft terms appearing for all SM particles, even if for some reason all Planck
mass supressed couplings of hidden sector fields to the SM were absent. They
are due to a conformal anomaly which arises because potentially
logarithmically divergent radiative corrections depend on a mass scale which
breaks conformal invariance. The atractive aspect of anomaly mediation is that
its effect is determined by the one-loop terms of the low-energy effective
theory, independent of ultraviolet physics, i.e.they only depend on the
$\beta$-functions and anomalous dimensions $\gamma_{f}$ of the SSM fields. In
particular one has
\begin{align}
{M_{a}}\  &  =\ \frac{{\beta_{a}}}{{g_{a}}}{m_{3/2}}\\
{m_{f}^{2}}\  &  =\ -\frac{1}{4}\left(  \frac{\partial{\gamma_{f}}}%
{\partial{g}}{\beta_{g}}+\frac{\partial{\gamma_{f}}}{\partial{Y}}{\beta_{Y}%
}\right)  {m_{3/2}^{2}}\\
{A_{Y}}\  &  =\ -\frac{{\beta_{Y}}}{{Y}}{m_{3/2}}%
\end{align}
wher $Y$ denote the corresponding Yukawa coupling. Note that soft terms are of
order $(\alpha/4\pi)m_{3/2}$ and gaugino masses are not universal but rather
are on the ratios $M_{1}:M_{2}:M_{3}=2.8:1:7.1$. This implies that typically
the LSP is the neutral wino and there is a relatively light chargino. As we
said the nicest feature of anomaly mediation is its independence of the UV
physics. It also leads to degenerate squarks masses, suppressing unwanted
FCNC. On the other hand, although this contribution is always present, it is
difficult to construct explicit supergravity/string scenarios in which they
are the leading effect (see however comments at the end of the next section).
Furthermore one finds from above formulae that the sleptons have negative
mass$^{2}$, which makes the simplest anomaly mediation models not viable. On
the other hand simple extensions including e.g. the contribution of D-terms
from an extra $U(1)$ symmetry\cite{Jones:2006re} or contributions from string
moduli\cite{Choi:2005ge} can cure this desease.

\subsection{String Theory and SUSY-breaking}

We have mentioned above several different possibilities considered in the
literature for the understanding of the origin and/or structure of SUSY
breaking in SUSY versions of the SM. Whatever the solution proposed, one would
like to embed the SUSY-breaking inside an ultraviolet consistent theory which
also incorporates gravity. The only candidate known to date for a consistent
unification of quantum gravity and the SM is string theory. One SUSY breaking
mechanism which appears very naturally in the context of string theory is
gravity mediation. Indeed, generic string compactifications with SM-like
massless quark/lepton spectrum have additional massless singlet chiral fields,
the \textit{moduli}, whose couplings to ordinary matter are Planck-mass
suppressed. These include the complex dilaton $S$ (whose real part is related
in some compactifications to the inverse gauge coupling constant), the Kahler
moduli $T_{i}$ (whose real parts describe the size of the compact 6 extra
dimensions) and the complex structure $U_{j}$ (which describe the shape of the
extra dimensions). These singlet fields are natural candidates to constitute
the hidden sector responsible for SUSY-breaking. Thus the idea is that it is
the non-vanishing of the auxiliary fields $F_{S}$,$F_{T_{i}}$, $F_{U_{i}}$
which would be the seed of SUSY breaking. As we will emphasize below, it has
been recently realized that non-vanishing values for certain 10-dimensional
bosonic fields (Ramond-Ramond and NS-NS backgrounds) do indeed give rise to
non-vanishing expectation values to the auxiliary fields of the moduli. Such
terms in turn generate SUSY-breaking soft terms for the SM fields. Another
ingredient which naturally appears in the context of string theory is the
generic presence of extra hidden sector gauge groups which may trigger
SUSY-breaking (inducing vevs for the auxiliary fields mentioned above) upon
gaugino condensation \cite{gc}.

In order to compute SUSY-breaking soft terms within this scheme, we need to
know the dependence of the gauge kinetic functions $f_{a}$ , the Kahler
metrics ${\tilde{K}}_{IJ}(S,T_{i},U_{j})$ of the SM chiral superfields and the
Kahler potential of the moduli $K(S,T_{i},U_{j})$. All these quantities may be
computed to leading order for some simple compactifications. Then gaugino
masses, scalar masses and trilinear A-terms may be computed from the general
expressions in eq.(\ref{softt}). The dependence of $f_{a}$, ${\tilde{K}}%
_{IJ}(S,T_{i},U_{j})$ and $K(S,T_{i},U_{j})$ on the moduli depend on the
specific way in which the SM is embedded into string theory. In particular
they depend on whether we are dealing with heterotic or e.g. Type IIA or Type
IIB orientifold D-brane models. This is not the place to give a detail review
of these issues but let us give some example of the kind of relationships
among soft terms that one can encounter in some simple compactification scenarios:

\begin{itemize}
\item \textit{Dilaton domination boundary conditions}. If the origin of SUSY
breaking in an heterotic string model is the auxiliary field of the dilaton
one finds the relationship among SUSY-breaking soft terms \cite{kl,bim}:
\begin{equation}
M\ =\ \sqrt{3}m\ =\ -A
\end{equation}
These boundary conditions are flavour independent, which is welcome in order
to avoid large FCNC effects. It turns out that such type of boundary
conditions do appear in other string constructions. For example, if the SM is
embedded inside (anti-)D3-branes in Type IIB orbifolds and/or orientifolds
\cite{ciu1}. In this case the origin of the non-vanishing auxiliary field for
the dilaton may be understood in terms of RR-NS flux backgrounds. These
boundary conditions are simple and very constraining. In fact if one takes at
face value these boundary conditions and does the renormalisation group
running from a GUT scale of order $10^{16}$ GeV down to low energies one finds
that charge- and color-breaking minima appear \cite{clm}. On the other hand,
as emphasized in \cite{aaqik} some string models with an intermediate string
scale $M_{string}\propto10^{10}$ GeV such charge/color-breaking minima disappear.

\item \textit{Heterotic-like T-modulus dominance}.

In heterotic compactifications if the auxiliary field of the volume modulus
$T$ is the only non-vanishing, one gets the standard no-scale structure
\cite{noscale}. This means in fact that to leading order all soft terms vanish
$M=m=A=0$. However both loop corrections and world-sheet instanton corrections
induce non-vanishing soft terms (see e.g.\cite{bim}). However those are model
dependent and no sharp model-independent prediction can be made.

\item \textit{Type IIB orientifold T-modulus dominance}. It turns out that
T-modulus dominance has a different effect on different types of string theory
vacua. In particular, in the case of Type IIB orientifolds with the SM gauge
group residing on D7-branes, if the SM fields are assumed to behave like the
scalars which parametrize the position of D7-branes in extra dimensions, one
obtains the simple boundary conditions \cite{fluxed}
\begin{equation}
m\ =\ M\ ;\ A\ =\ -3M\ ;\ B\ =\ 2M\mu
\end{equation}
This scheme, called `fluxed MSSM' in ref.\cite{fluxed} is very restrictive
since it also gives a prediction for the elusive $B$-parameter and has no SUSY
CP-problem. In ref.\cite{abi} it was shown that these relationships are
consistent with correct radiative electroweak symmetry breaking and other
phenomenological constraints. If one insists in obtaining the appropriate
neutralino dark matter abundance a quite heavy spectrum of sparticles is
predicted \cite{abi}.

\item \textit{Modulus dominance in a specific MSSM-like IIB orientifold}. One
can construct concrete Type IIB orientifold models with overlapping
`magnetized' D7-branes with a massless spectrum very close to that of the MSSM
\cite{guay}. These are models based on Type IIB string theory compactified on
a $Z_{2}\times Z_{2}$ orientifold with D7-branes. In this settings one can
compute the relevant $f_{a}$, ${\tilde{K}}_{IJ}(S,T_{i},U_{j})$ and
$K(S,T_{i},U_{j})$ to leading order \cite{imrmas,lrs2}. Assuming that the
auxiliary field of the overall modulus T dominates one finds in a certain
approximate limit soft terms of the form \cite{fi}
\begin{equation}
m^{2}\ =\ \frac{M^{2}}{2}\ ;\ A\ =\ -\frac{3}{2}M\
\end{equation}
Again in these examples one can understand the non-vanishing value of the
modulus auxiliary field $F_{T}$ as originating on the presence of certain RR
and NS 10-dimensional field fluxes. A study of the phenomenological
consequences of analogous models has been presented in \cite{kklw}.
\end{itemize}

These examples do not exhaust at all the possibilities but they show that, at
least in certain simplified string model compactifications, one can obtain
specific predictions for the structure of soft terms. Other schemes have been
considered recently in the context of variants of the KKLT scenario
\cite{kklt} for moduli fixing in string theory. In particular in
refs.\cite{ccq} it has been shown that in certain KKLT-like scenarios of
moduli fixing with large-volume Calabi-Yau flux compactifications of Type IIB
string theory, one recovers in certain approximation the dilaton dominated
kind of boundary conditions described above. In other KKLT-inspired schemes it
has been argued that a mixture of anomaly and modulus mediation arises
\cite{cfno}. Recent progress has also been achieved in obtaining string models
with gauge mediated SUSY-breaking \cite{gaugemed}. In any event it is clear
that, if SUSY is found at LHC, measuring the SUSY spectrum and couplings would
give important information about the structure of the underlying string theory.

\section{Radiative Electroweak Symmetry Breaking}


Once one has an understanding of the soft SUSY\ breaking parameters it is
possible to address the question whether there is a stage of spontaneous
symmetry breaking by studying the scalar potential. As we have seen in the
supersymmetric extension of the Standard Model there are a large number of
scalar fields and many possible direction of spontaneous symmetry breaking
corresponding to the various possible combinations of these scalar fields that
can acquire vacuum expectation values (vevs). To simplify the discussion we
first discuss the potential involving only the Higgs scalar fields and later
return to a discussion why these are the only fields that acquire vevs,
explaining why the breaking is in the direction $SU(3)\times SU(2)\times U(1)$
$\rightarrow SU(3)\times U(1).$

From the gauge interactions and the interactions following from the
superpotential of eq(4) the Higgs scalar potential has the form%

\begin{align}
V(H_{1},H_{2})\  &  =\ {\frac{{{g_{2}}^{2}}}{2}}\ (H_{1}^{\ast}{\frac
{{\tau^{a}}}{2}}H_{1}+H_{2}^{\ast}{\frac{{\tau^{a}}}{2}}H_{2})^{2}%
\ +\ {\frac{{{g_{1}}^{2}}}{8}}\ (|H_{1}|^{2}-|H_{1}|^{2})^{2}\ \nonumber\\
&  +\ {\mu_{1}}^{2}\ |H_{1}|^{2}\ +\ {\mu_{2}}^{2}\ |H_{2}|^{2}\ -\ {\mu_{3}%
}^{2}(H_{1}H_{2}\ +h.c.\ )\label{pot}%
\end{align}
where $\tau^{a},a=1,2,3$ are the $SU(2)_{L}$ Pauli matrices and
\begin{equation}
{\mu_{1}}^{2}\ \equiv m_{0}^{2}\ +\ \mu_{0}^{2}\ ;\ {\mu_{2}}^{2}%
\ \equiv\ m_{0}^{2}\ +\ \mu_{0}^{2}\ ;\ {\mu_{3}}^{2}\ \equiv\ B_{0}\mu
_{0}\ .\label{bounda}%
\end{equation}
This is the SUSY version of the `Mexican hat' Higgs potential of the standard
model. However, this potential as it stands looks problematic. Indeed, in
order to get a non-trivial minimum we need to have a negative $mass^{2}$
eigenvalue in the Higgs mass matrix, i.e., we need $\mu_{1}^{2}\mu_{2}^{2}%
-\mu_{3}^{4}\ <0$. However, since $\mu_{1}^{2}=\mu_{2}^{2}>0$, this may only
happen if $\mu_{1}^{4}=\mu_{2}^{4}<\mu_{3}^{4}$ in which case the scalar
potential is unbounded below in the direction $<H_{1}>=<H_{2}>\rightarrow
\infty$. The puzzle is resolved \cite{Ibanez:1982fr},\cite{Inoue:1982pi}%
,\cite{Alvarez-Gaume:1983gj},\cite{irb} \ by noting that the boundary
conditions eq.\ref{bounda}\ apply only at the unification or Planck scale. At
any scale below one has to consider the quantum corrections to the scalar
potential which can be substantial. Consider for example the one-loop
corrections to the masses of the Higgs fields. These come from graphs which
involve the Yukawa couplings of the Higgs field $H_{2}$ to the u-type quarks
and squarks. Of course, these corrections will be negligible except for the
ones involving the top quark which has a relatively large Yukawa coupling (for
simplicity we ignore here the possibility of a large bottom Yukawa coupling).
\begin{figure}
[ptb]
\begin{center}
\includegraphics[
height=1.6457in,
width=4.5161in
]%
{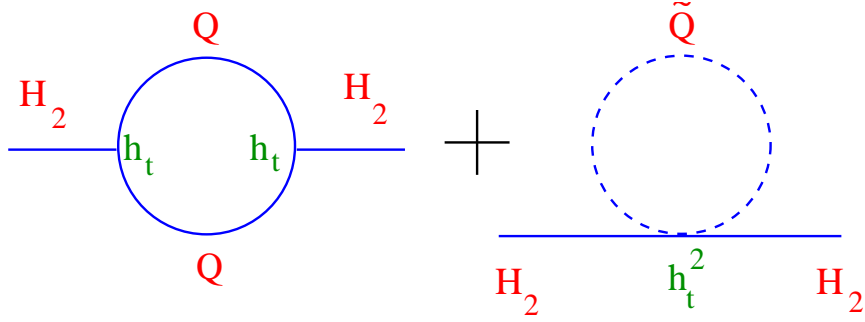}%
\caption{One loop contributions to the Higgs mass in the MSSM.}%
\label{fig1}%
\end{center}
\end{figure}
While supersymmetry is a good symmetry the first graph in fig. 1 leads to a
(negative) quadratically divergent contribution which is exactly cancelled by
the second graph. Once susy is broken the sparticles get masses and the
right-hand diagram is suppressed compared to the left-hand one leaving an
overall uncanceled $negative$ contribution \cite{Ibanez:1982fr}%
,\cite{Inoue:1982pi},\cite{Alvarez-Gaume:1983gj},\cite{irb}%
\begin{equation}
\delta\mu_{2}^{2}\ \simeq\ -{\frac{3}{{16\pi^{2}}}}h_{t}^{2}m_{\tilde{Q}}%
^{2}\log(M_{X}^{2}/(\mu^{2}+m_{\tilde{Q}}^{2}))\ .\label{semin}%
\end{equation}
where the contribution is evaluated at a scale $\mu.$ If $h_{t}$ is large
enough (i.e., if the top quark is heavy enough) this negative contribution
may, at a scale below M$_{X},$ overwhelm the original positive contribution
and trigger electroweak symmetry breaking. Similar diagrams exist for the
other Higgs field $H_{1}$ but those are expected to give small contributions
since they will be proportional to the bottom Yukawa coupling.%

\begin{figure}
[ptb]
\begin{center}
\includegraphics[
height=1.8706in,
width=3.6824in
]%
{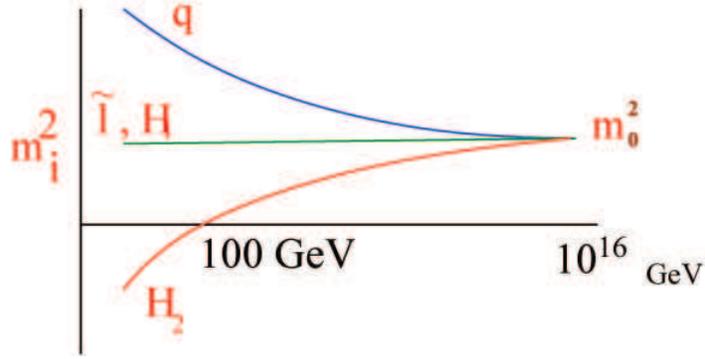}%
\caption{Loop effects trigger the electroweak symmetry breaking in a natural
way. The mass$^{2}$ of the Higgs scalars gets negative as the energy
decreases, meanwhile squark/slepton mass$^{2}$ remain positive.}%
\label{radiativesb1}%
\end{center}
\end{figure}

One may worry that one can find similar graphs involving squarks such as
${\tilde{t}},{\tilde{b}}$ that could drive their $mass^{2}$ negative leading
to minima with broken charge and colour. Indeed these graphs exist but
coloured scalars also get large ({\ positive}) contributions to their
$mass^{2}$ from loops involving gluinos which are proportional to the large
strong coupling constant, preventing $SU(3)\otimes U(1)_{em}$ breaking. The
resulting pattern of running masses for the Higgs scalar and the various
sparticles of the MSSM is shown in Figure 2. From this one sees that the
structure of the minimal supersymmetric standard model is such that quantum
corrections select the desired pattern of $SU(3)\otimes SU(2)\otimes U(1)$
symmetry breaking in a natural and elegant way. Note that the proposal for
radiative electroweak breaking relied on a heavy top quark anticipating the
subsequent measurement of the top quark mass, $m_{t}=$ $O(170GeV)!$

With $\mu_{2}^{2}\not =\mu_{1}^{2}$ the potential in eq.(\ref{pot}) is
perfectly well behaved and one can see it is minimized for
\cite{Inoue:1982pi,INOU}%
\begin{equation}
\nu^{2}\ \equiv\ \nu_{1}^{2}+\nu_{2}^{2}\ =\ {\frac{{2(\mu_{1}^{2}-\mu_{2}%
^{2}-(\mu_{1}^{2}+\mu_{2}^{2})cos2{\beta})}}{{(g_{2}^{2}+g_{1}^{2})cos2\beta}%
}} \label{min}%
\end{equation}
where $\nu_{1,2}=<H_{1,2}^{0}>$ and $sin2\beta\equiv2\mu_{3}^{2}/(\mu_{1}%
^{2}+\mu_{2}^{2})$. The existence of a non-vanishing $\mu_{3}^{2}$ forces the
two vevs to be aligned in such a way that electric charge remains unbroken.
The W mass is related to $\nu$ via $\nu^{2}=2M_{W}^{2}/g^{2}$. This condition
may be equivalently written \cite{IBLO}%

\begin{equation}
{\frac{{\nu_{2}^{2}}}{{\nu_{1}^{2}}}}\ =\ {\frac{{\mu_{1}^{2}+{\frac{1}{2}%
}M_{Z}^{2}}}{{\mu_{2}^{2}+{\frac{1}{2}}M_{Z}^{2}}}}\ \label{const}%
\end{equation}
where $\mu_{1,2}^{2}$ should be evaluated at the weak scale. Thus in a model
with the correct $SU(2)\times U(1)$ breaking the parameters are constrained in
such a way that $\mu_{1,2}^{2}(M_{W})$ and $\mu_{3}^{2}(M_{W})$ satisfy the
above conditions. As we discussed in Section 2 the free parameters in the
minimal supergravity model are just%

\begin{equation}
m_{0}\ ,\ m_{1/2}\ ,\ A_{0}\ ,\ \mu_{03}^{2}\equiv B_{0}\mu_{0}\ ,\ \mu_{0}
\label{param}%
\end{equation}
plus the Yukawa couplings, of which $h_{t}$ is likely to be the only one
playing an important role in the running of the soft terms. In order to see
how all these parameters are constrained we need to use the renormalisation
group equations which relate the values of couplings and masses at the
unification scale with their values at the weak scale. The renormalisation
group equations generalise the analysis of radiative corrections given in
eq.\ref{semin} allowing for the summation of all powers of the logarithmic corrections.

\subsection{Renormalisation Group analysis}

The renormalisation group equations for the Yukawa couplings \cite{INOU}\ can
be integrated analytically in the case in which one only keeps the top-quark
Yukawa coupling. For the third generation one finds \cite{IBLO}%
\begin{align}
h_{t}^{2}(t)\  &  =\ h_{t}^{2}(0)\ {\frac{{E_{1}(t)}}{{1+6Y_{t}(0)F_{1}(t)}}%
}\\
h_{b}^{2}(t)\  &  =\ h_{b}^{2}(0)\ {\frac{{E_{2}(t)}}{{(1+6Y_{t}%
(0)F_{1}(t))^{1/6}}}}\\
h_{\tau}^{2}(t)\  &  =\ h_{\tau}^{2}(0)\ E_{3}(t)\ , \label{renuk}%
\end{align}
where $t\equiv2\log(M_{X}/Q)$ and $E_{1,2,3}$ and $F_{1}$ are known functions
of $\alpha_{i}(t)$ and $Q$ is the scale at which the couplings are evaluated.
The $E_{i}$ functions give just the usual gauge anomalous dimension
enhancement whereas the effect of the top Yukawa coupling in the running gives
the extra factor. Let us first discuss the case of the top quark. Notice that
for small $h_{t}(0)$ one recovers the well-known gauge anomalous dimension
result. However, for $Y_{t}(0)\rightarrow\infty$ one gets
\begin{equation}
h_{t}^{2}(t)\ =\ {\frac{{(4\pi)^{2}E_{1}(t)}}{{6F_{1}(t)}}} \label{htop}%
\end{equation}
independently of the original value of $Y_{t}(0)$, i.e., there is an infrared
fixed point. At the weak scale ($t\simeq67$) one obtains $E_{1}\simeq13$ and
$F_{1}\simeq290$ which gives an upper bound for the top-quark mass
\begin{equation}
m_{t}\ =\ h_{t}\nu_{2}\ \leq\ h_{t}\nu\ \leq\ 190\ GeV\ .
\end{equation}
Let us now consider the running of the mass parameters which are the ones of
direct relevance to the $SU(2)\times U(1)$-breaking process. In particular,
consider the running of the squarks, sleptons and Higgs masses. The
renormalisation group equations describing the mass$^{2}$ evolution have a
gauge contribution proportional to gaugino masses and a second contribution
proportional to the top-Yukawa coupling$^{2}$. The gauge piece makes the
mass$^{2}$ increase as the energy decreases. In particular, squarks get more
and more massive as we go to low energies since their equation is proportional
to $\alpha_{3}$. The piece in the equations proportional to the top Yukawa
coupling has the opposite effect and decreases the mass$^{2}$ as the scale
decreases. This effect is normally not big enough to overwhelm the large
positive contribution to the mass$^{2}$ of squarks involving the QCD coupling
but may be sufficiently large to overwhelm the positive contribution of weakly
interacting scalars which only involve the electroweak couplings. The only
such scalar in which this can happen is $H_{2}$ since it is the only one
(unlike sleptons) which couples directly to the top Yukawa. This is nothing
but the renormalisation group improved version of the mechanism in
eq.\ref{semin}\ . We thus see that the quantum structure of the MSSM leads
automatically to the desired pattern of symmetry breaking in a natural way.
The qualitative behaviour of the running of scalars is shown in fig. 2.

Apart of obtaining the desired pattern of symmetry breaking one is interested
in finding out the spectrum of sparticles in this scheme. Since there are only
a few free parameters one has strong predictive power. In the case of the
squarks and sleptons, integration of the renormalisation group equations
(neglecting Yukawa couplings) leads to the following result \cite{IBLO}%
,\cite{Martin:1997ns}%
\begin{align}
m_{{\tilde{U}}_{L}}^{2}\  &  =\ m_{0}^{2}+2m_{1/2}^{2}({\frac{4}{3}}%
{\tilde{\alpha}}_{3}f_{3}+{\frac{3}{4}}{\tilde{\alpha}}_{2}f_{2}+{\frac
{1}{{36}}}{\tilde{\alpha}}_{1}f_{1})+cos(2\beta)M_{Z}^{2}({\frac{{-1}}{2}%
}+{\frac{2}{3}}sin^{2}\theta_{W})\\
m_{{\tilde{D}}_{L}}^{2}\  &  =\ m_{0}^{2}+2m_{1/2}^{2}({\frac{4}{3}}%
{\tilde{\alpha}}_{3}f_{3}+{\frac{3}{4}}{\tilde{\alpha}}_{2}f_{2}+{\frac
{1}{{36}}}{\tilde{\alpha}}_{1}f_{1})+cos(2\beta)M_{Z}^{2}({\frac{1}{2}}%
-{\frac{1}{3}}sin^{2}\theta_{W})\nonumber\\
m_{{\tilde{U}}_{R}}^{2}\  &  =\ m_{0}^{2}+2m_{1/2}^{2}({\frac{4}{3}}%
{\tilde{\alpha}}_{3}f_{3}+{\frac{4}{9}}{\tilde{\alpha}}_{1}f_{1}%
)-cos(2\beta)M_{Z}^{2}({\frac{2}{3}}sin^{2}\theta_{W})\nonumber\\
m_{{\tilde{D}}_{R}}^{2}\  &  =\ m_{0}^{2}+2m_{1/2}^{2}({\frac{4}{3}}%
{\tilde{\alpha}}_{3}f_{3}+{\frac{1}{9}}{\tilde{\alpha}}_{1})+cos(2\beta
)M_{Z}^{2}({\frac{1}{3}}sin^{2}\theta_{W})\nonumber\\
m_{{\tilde{E}}_{L}}^{2}\  &  =\ m_{0}^{2}+2m_{1/2}^{2}({\frac{3}{4}}%
{\tilde{\alpha}}_{2}f_{2}+{\frac{1}{4}}{\tilde{\alpha}}_{1}f_{1}%
)+cos(2\beta)M_{Z}^{2}({\frac{1}{2}}-sin^{2}\theta_{W})\nonumber\\
m_{{\tilde{\nu}}_{L}}^{2}\  &  =\ m_{0}^{2}+2m_{1/2}^{2}({\frac{3}{4}}%
{\tilde{\alpha}}_{2}f_{2}+{\frac{1}{4}}{\tilde{\alpha}}_{1}f_{1}%
)-cos(2\beta){\frac{1}{2}}M_{Z}^{2}\nonumber\\
m_{{\tilde{E}}_{R}}^{2}\  &  =\ m_{0}^{2}+2m_{1/2}^{2}({\tilde{\alpha}}%
_{1}f_{1})+cos(2\beta)M_{Z}^{2}sin^{2}\theta_{W} \nonumber\label{masillas}%
\end{align}
where $\theta_{W}$ is the weak angle, $M_{Z}$ is the $Z^{0}$ mass and
$m_{0},m_{1/2}$ and $tg\beta=\nu_{2}/\nu_{1}$ are related to the free
parameters in eq.\ref{param}. In this equation
\begin{equation}
{\tilde{\alpha}}_{i}\ \equiv\ {\frac{{\alpha_{i}(M_{X})}}{{(4\pi)}}}%
\ ;\ f_{i}\ \equiv\ {\frac{{(2+b_{i}{\tilde{\alpha}}_{i}t)}}{{(1+b_{i}%
{\tilde{\alpha}}_{i}t)^{2}}}}t \label{auxil}%
\end{equation}
where $b_{i}=(-3,1,11)$ are the one-loop coefficients of the $\beta$-function
of the $SU(3)\otimes SU(2)\otimes U(1)$ interactions. The above equations
assume universal soft masses $m_{0}$ for all the scalars in the theory at the
unification scale as well as universal gaugino masses $m_{1/2}$. The
right-most term in eqs.\ref{masillas}\ does not in fact come from the
integration of the r.g.e.'s but from the contribution of the D$^{2}$-term in
the scalar potential of sfermions once $SU(2)\times U(1)$ is broken. As noted
above the squarks will be heavier than the sleptons since $\alpha_{3}\gg
\alpha_{2}$.

\section{The Little Hierarchy Problem}

Low energy supersymmetry was introduced to solve the hierarchy problem. To
quantify how well it has achieved this we need to modify eq(\ref{hierarchysm})
to include the contributions of the new SUSY\ states as in eq(\ref{semin}).
Solving for the minimum of the Higgs scalar potential one finds a bound on the
Higgs mass of the form \cite{higgssloop}%

\begin{equation}
m_{h}^{2}\leq M_{Z}^{2}\cos^{2}2\beta+{\frac{3m_{t}^{4}}{2\pi^{2}v^{2}}}%
\log{\frac{M_{\widetilde{\mathrm{t}}}^{2}}{m_{t}^{2}}}+... \label{mh}%
\end{equation}
where $M_{\widetilde{\mathrm{t}}}^{2}$ is an average of stop masses squared.
The first term is the tree level contribution to the Higgs mass, small because
in SUSY\ the quartic Higgs coupling, c.f. eq(\ref{one}) is given by
$\lambda=(1/8)(g^{2}+g^{\prime2})\cos^{2}2\beta$ . The radiative corrections
are necessary to increase $m_{h}$ beyond the experimental bound, $m_{h}%
\geq115$ GeV. In order to satisfy this mass bound one is forced to live in a
region of relatively large soft masses, $M_{\widetilde{\mathrm{t}}}\geq300$ GeV.

This leads to a conflict with the hierarchy problem because the mass is so
large that it requires some fine tuning to keep the electroweak breaking scale
low. To see this note that the electroweak breaking scale, which is
characterised by $M_{Z},$ depends on the input parameters of the theory. The
input parameters are defined at the SUSY breaking messenger scale, $M_{mess},
$ and the RG running can introduce large logs if the scale is large which in
turn drive the electroweak breaking scale large. The $Z$ mass immediately
follows from eq(\ref{min}) and the main contribution to the negative Higgs
mass squared parameter triggering electroweak breaking coming from loops of
tops and stops is approximately given by
\[
-\frac{3h_{t}^{2}}{4\pi^{2}}(M_{\widetilde{\mathrm{t}}}^{2}+|A_{t}|^{2}%
)\ln\left(  \frac{M_{mess}}{M_{\widetilde{\mathrm{t}}}}\right)  .
\]
Putting all this together in the case of the MSSM with gravity mediation,
$M_{mess}=M_{Planck},$ one finds for $\tan\beta=2.5$\cite{Kane:1998im}%

\begin{align}
\frac{M_{Z}^{2}}{2}  &  =-.87\,\mu^{2}(0)+3.6\,{M_{3}^{2}(0)}-.12\,{M_{2}%
^{2}(0)}+.007\,{M_{1}^{2}(0)}\nonumber\\
&  -.71\,{m_{H_{U}}^{2}(0)}+.19\,{m_{H_{D}}^{2}(0)}+.48\,({m_{Q}^{2}%
(0)}+\,{m_{U}^{2}(0)})\nonumber\\
&  -.34\,{A_{t}(0)}\,{M_{3}(0)}-.07\,{A_{t}(0)}\,{M_{2}(0)}-.01\,{A_{t}%
(0)}\,{M_{1}(0)}+.09\,{A_{t}^{2}(0)}\nonumber\\
&  +.25\,{M_{2}(0)}\,{M_{3}(0)}+.03\,{M_{1}(0)}\,{M_{3}(0)}+.007\,{M_{1}%
(0)}\,{M_{2}(0)} \label{MZ}%
\end{align}
where the coefficients reflect the effect of the large logs in the running.
For the soft mass parameters of $O(300GeV)$ one sees from this formula that
the measured value of $M_{Z\text{ }}$ can only be achieved through a
cancellation of terms much larger than $M_{Z}.$ To quantify this Barbieri and
Giudice \cite{Barbieri:1987fn} defined a measure of fine tuning, $\Delta_{a},$
as the fractional change in the Z mass squared per unit fractional change in
the input parameter,
\begin{equation}
\Delta_{a}=Abs\left(  \frac{a}{M_{Z}^{2}}\frac{\partial M_{Z}^{2}}{\partial
a}\right)
\end{equation}
for each input parameter $a$.

%

\begin{figure}
[ptb]
\begin{center}
\includegraphics[
height=3.5405in,
width=3.2586in
]%
{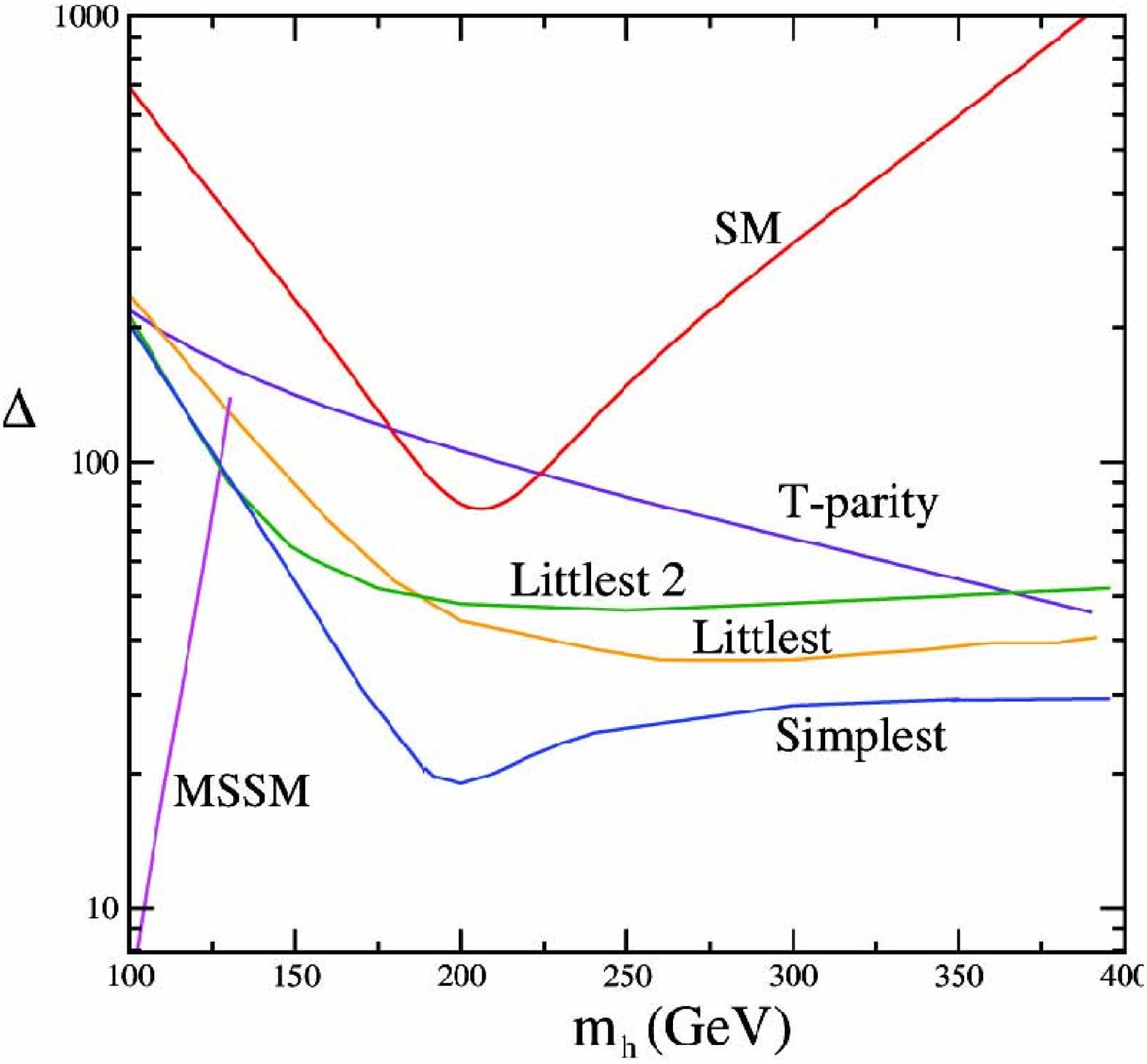}%
\caption{The finetuning measure for a variety of models \cite{Casas:2005ev}.}%
\label{casasfig}%
\end{center}
\end{figure}

Due to the quadratic divergence noted in eq(\ref{hierarchysm}) in the Standard
Model, with a cutoff at the GUT scale, the fine tuning measure is
$\Delta=O(10^{30})!$ This suggests that new physics is needed at a low scale
but the nature of the new physics is limited by the need not to conflict with
the successful precision tests of the Standard Model . Treating the Standard
Model as an effective field theory below the cut-off scale there will be
higher dimension operators involving the Standard Model fields suppressed by
powers of the cut-off scale, $\Lambda.$ Assuming these operators appear
unsuppressed, the minimum scale consistent with the precision tests is
$\Lambda=10TeV.$ Using this cut-off one may calculate the residual fine tuning
measure $\Delta=\left(  \sum_{a}\Delta_{a}^{2}\right)  ^{1/2}$ in the Standard
Model as a function of the Higgs mass. This is shown in Figure \ref{casasfig}%
\cite{Casas:2005ev} by the curve labeled $SM,$ the dominant effect coming from
the contribution to the Higgs mass from the top quark loop, and has
$\Delta\geqslant80.$ This residual need for fine tuning even if physics beyond
the Standard Model is allowed is the \textquotedblleft little hierarchy
problem\textquotedblright.

To do better some of the new operators must be suppressed and this requires
some symmetry to organise the corrections to the electroweak breaking scale so
that they largely cancel. In the $MSSM$ supersymmetry plays this role and its
fine tuning measure is also shown in Figure \ref{casasfig}. One may see that
$\Delta\geqslant20$, the lowest tuning only applying for a Higgs close to its
current mass limit. This is a significant improvement over the naive estimate
using the low cut-off Standard Model\footnote{The reason supersymmetry reduces
this residual fine tuning is through cancellation of the top contribution with
the new supersymmetric contributions, particularly that of the top squark.}
and certainly a huge improvement over the GUT\ scale cut-off. Nonetheless the
fact that $\Delta$ is still much larger than $1$ is worrying and much work has
been done exploring ways of reducing the residual fine tuning. As we have seen
the origin of the little hierarchy problem is the tension between the need to
have the Higgs heavy, eq(\ref{mh}) and the need to keep the $Z$ light,
eq(\ref{MZ}). The latter equation follows from the radiative electroweak
breaking c.f. eq(\ref{min}) and so the little hierarchy problem raises doubts
on the validity of this mechanism. For this reason we think it appropriate
here to discuss briefly the attempts to eliminate this residual fine tuning.

In a supersymmetric context the residual fine tuning may be due to the choice
of soft supersymmetry breaking parameters rather than the underlying
supersymmetric theory itself. One may see from eq(\ref{MZ}) that the gluino
contribution proportional to $\,{M_{3}^{2}(0)}$ is particularly large.
Reducing it reduces the fine tuning and this can be done in several ways. One
can take a non-GUT symmetric form of the gaugino masses, reducing ${M_{3}%
^{2}(0)}$ relative to ${M_{1,2}^{2}(0).}$ Alternatively one can lower the
scale at which the soft mass terms are defined (the messenger scale) thus
reducing the magnitude of the logarithmic radiative terms responsible for the
large coefficient of ${M_{3}^{2}(0)}$ and also adding new contributions to the
quartic Higgs coupling${.}$ These changes can reduce the fine tuning needed,
$\Delta\geqslant10$ $\cite{Casas:2004uu},\cite{Chacko:2005ra}$ and even
eliminate it in regions $\ $of parameter space with $M_{\widetilde{\mathrm{t}%
}}$ small and $A_{t}$ large\cite{Kitano:2005wc}. In schemes with anomaly
mediation, although the initial spectrum is different, the fine tuning
remains. An example of this called "mirage unification" in which new SUSY
breaking contributions from moduli solve the tachyonic slepton problem, has
been studied in detail. Although it does reduce the gluino mass in simple
versions of the model there is still significant fine tuning needed,
$\Delta\geqslant10^{3}$ $\cite{Lebedev:2005ge}.$ Another possibility, still in
the context of supersymmetry, is to consider a more complicated scalar sector.
This can affect the fine tuning needed, either by making the lightest Higgs
invisible thus reducing the constraint on $M_{\mathrm{SUSY}}$ from the Higgs
mass bounds, c.f. eq(\ref{mh}), or by adding new contributions to the quartic
Higgs coupling which increase the tree level contribution to eq(\ref{mh}) and
again reduce the constraint on $M_{\mathrm{SUSY}}.$ Again this only marginally
improves the situation, $\Delta\geqslant10\cite{Schuster:2005py}.$

What about the non-supersymmetric attempts to eliminate the little hierarchy
problem? To do this it is necessary to have new contributions beyond the
Standard Model ordered by a symmetry that prevent the radiative corrections to
the Higgs mass becoming large, thus cancelling in part the top quark loop
contribution. The only symmetry, apart from supersymmetry, capable of doing
this is a (pseudo)Goldstone symmetry in which the components of the Standard
Model Higgs doublet are Goldstone bosons associated with an underlying
symmetry. It is known that the Standard Model gauge couplings do not respect
such a symmetry and must give mass to the physical Higgs. However if the
Goldstone symmetry is still unbroken at the level of the quadratically
divergent radiative corrections there will be no Higgs mass corrections of the
form of eq(\ref{hierarchysm}). In the little Higgs model \cite{lh} this is
arranged to happen at one loop order, the leading radiative corrections to the
Higgs mass occurring at logarithmic and finite order. To keep these latter
contributions under control it is necessary that there are new physics
contributions coming from states much lighter than the $10TeV$ cutoff assumed
when calculating the fine tuning measure\footnote{This cutoff is needed to
control the quadratic divergences at two loop order.} and beyond which some
unspecified ultra-violet completion is assumed. The new physics contributions
are needed to cancel the (irreducible) contribution of the top quark loop. In
Figure \ref{casasfig} we show the resultant fine tuning for four
representative little Higgs models (see reference \cite{Casas:2005ev} for
details). In the littlest Higgs model the top contribution is largely
cancelled by the contribution of a new "top" quark, vector-like with respect
to the Standard Model gauge group with a mass around $1TeV.$ One may see that
there is still significant fine tuning needed, $\Delta\geqslant40,$ with the
lower values occurring only for a heavy Higgs which is difficult to reconcile
with the precision tests of the Standard Model. The second little Higgs model
and the $T-$parity little Higgs models are quite similar to the littlest Higgs
model with small changes to the heavy spectrum and to the allowed couplings.
As may be seen from the figure they do not change the fine tuning very much.
The littlest Higgs model starts with a different symmetry group structure but
still has a heavy top. For a heavy Higgs it improves on the fine tuning
needed, $\Delta\geqslant20,$ but does not eliminate the need for fine tuning.
It is possible to build models protected by both a (pseudo)Goldstone symmetry
and supersymmetry. Such models have double protection against large radiative
corrections and can eliminate the fine tuning
completely\cite{Falkowski:2006qq}. However the models are very complicated and
it is hard to believe nature would go to such lengths to hide supersymmetry
from us!

Given all this what conclusions can we draw from the little hierarchy problem?
The first is that the non-supersymmetric attempts so far examined do
\textit{not} improve on the fine tuning needed even when comparing to the
simplest MSSM scheme with a low Higgs mass. Moreover they do not attempt to
provide an ultraviolet completion and thus do not address the question of a
possible underlying unification of forces. On the other hand models with
supersymmetry do provide an ultraviolet completion which allows for a stage of
gauge or string unification at a high scale while separating the electroweak
scale from the unification scale. For a low value of the Higgs mass the
residual fine tuning needed is relatively modest, even in the simplest
supersymmetric extensions of the Standard Model. There could also be
relationships among the fundamental soft parameters in such a way that they
should not be varied independently and hence the computed fine-tuning could be
an overestimation (see e.g. the case with $\mu= -2M$ in ref.\cite{abi}). Given
that these SM extensions lead to a very attractive picture of the theory at
high scales in which the gauge couplings and soft masses unify and radiative
electroweak breaking naturally explains the pattern of symmetry breaking in
the Standard Model, we consider them to be very good candidates for extensions
of the Standard Model. As we have discussed the residual fine tuning is
sensitive to the soft supersymmetry breaking parameters and some reduction of
fine tuning, even at higher Higgs mass, is possible without losing all the
benefits of an underlying unification. As such they do provide an answer to
the questions posed by the hierarchy problem and we think avoid the need to
turn to more drastic ideas involving anthropic arguments\cite{anthropics}, at
least in what concerns the electroweak scale.

\section{Summary and Prospects for SUSY\ and Higgs searches.}

The need to solve the hierarchy problem is a strong motivation for low energy
supersymmetry. Indeed the need to control radiative corrections means that
supersymmetry is an essential ingredient of any viable string or GUT
unification which unifies the gauge couplings at a very high scale. The fact
that in the minimal supersymmetric extension of the Standard Model the
prediction for gauge coupling unification is good to better than $1\%$
accuracy lends strong support to such supersymmetric unification. To complete
the theory it is necessary to add soft supersymmetry breaking terms which must
originate from a hidden sector. The radiative corrections that communicate the
supersymmetry breaking lead naturally to a subsequent stage of gauge symmetry
breaking at a low scale. Remarkably, for a heavy top quark, they naturally
explain the pattern of symmetry breaking observed in the Standard Model,
breaking the electroweak symmetry while leaving the symmetries associated with
the strong and electromagnetic interactions unbroken.

The scale for the new SUSY\ particles is limited by our wish to solve the
hierarchy problem to be in a range accessible to the LHC. The details of the
spectrum depend on the detailed mechanism responsible for supersymmetry
breaking. This in turn is limited by constraints from flavour changing neutral
currents, electroweak symmetry breaking, dark matter abundance etc. What is
interesting is that all these constraints \textit{can} be satisfied in very
simple supersymmetry breaking schemes and these schemes provide our "best bet"
for the physics Beyond the Standard Model. The archetypical model starts with
the MSSM which has a stable LSP candidate for dark matter. Provided there are
no additional gauge non singlet fields precision gauge unification is
preserved\footnote{Even if the additional multiplets come in complete
multiplets, at two loop order they disturb the precision
agreement\cite{ghilencea}, disfavouring gauge mediated of SUSY\ breaking.}
suggesting mSUGRA mediated SUSY breaking as described below eq(\ref{softt}).
The scheme has degenerate scalar masses, $m_{0},$ at a high scale so the
flavour changing neutral currents are under control. A scan of parameter space
shows that quite naturally the breaking of electroweak symmetry proceeds
radiatively. Furthermore the dark matter abundance of the LSP can, for a
limited parameter range, explain the dark matter abundance of the universe.

We find it remarkable that this simple favoured scheme is consistent with all
know data. Moreover it is encouraging that it will be testable in the near
future. Recent fits
\cite{Ellis:2005tu,Allanach:2006jc,Ellis:2006ix,deAustri:2006pe,Ellis:2006nx}
to the parameters favours a solution in which the lightest Higgs is very
light, with a distribution peaked around $M_{H}=115GeV$ and bounded by
$M_{H}<127GeV,$ as is favoured by the need to minimise the little hierarchy
problem, and close to the lower bound established by LEP and consistent with
the precision tests of the Standard Model. In addition the SUSY\ breaking
scale could be low corresponding to gluinos and other sparticles being readily
produced at the LHC. The prospects are good that the LHC will not only be able
to discover the Higgs, thus establishing the origin of mass, but also find
supersymmetric states and open the way to establishing what lies \ Beyond the
Standard Model.

\bigskip

This work has been partially supported by CICYT (Spain), the Comunidad de
Madrid (project HEPHACOS) and the European Commission under the RTN program MRTN-CT-2004-503369.

\end{document}